\input harvmac
\def\s{{\sigma}}
\def\apm{\alpha^\prime}

\lref\gs{S. Giddings and A. Strominger, Phys. Rev. D{\bf 46} (1992) 627,
 hep-th/9202004.}
\lref\verlinde{R. Dijkgraaf, E. Verlinde and H. Verlinde, 
{\it Notes on matrix and micro strings}, hep-th/9709107.}
\lref\cghs{C. Callan, S. Giddings, J. Harvey and A. Strominger, 
Phys. Rev. D{\bf 45},(1992) 1005, hep-th/9111056. }
\lref\ghas{G. Horowitz and A. Strominger, Nucl. Phys. B{\bf 360} (1991) 197.}
\lref\jmf{J. Maldacena, Nucl .Phys. B{\bf 477} (1996) 168, 
hep-th/9605016.  }
\lref\db{M. Douglas and M. Berkooz, Phys. Lett. B{\bf 395} (1997) 196,
hep-th/9610236. }
\lref\nte{N. Seiberg, {\it New theories in six-dimensions and
matrix theory description of M theory on $T^5$ and $T^5/Z_2$},
hep-th/9705221.}
\lref\ewch{E. Witten, {\it On the conformal field theory of the Higgs branch},
hep-th/9707093.}
\lref\mm{
O. Aharony, M. Berkooz, S. Kachru, N. Seiberg and E. Silverstein, 
{\it Matrix description of interacting theories in six dimensions},
hep-th/9707079.}
\lref\ewm{E. Witten, Phys. Rev. D{\bf 44} (1991) 314. }
\lref\sfs{T. Banks, W. Fischler, S. Shenker and L. Susskind, Phys. Rev.
D{\bf 55} (1997) 5112, hep-th/9610043.}
\lref\fiver{N. Seiberg, Phys. Lett. B {\bf 388} (1996) 753, hep-th/9608111; 
N. Seiberg and D. Morrison, Nucl. Phys. B{\bf 483} (1997) 
229, 
hep-th/9609070.  }
\lref\sixr{E. Witten, Proceedings of Strings 95, hep-th/9507121; 
 A. Strominger, Phys. Lett. B {\bf 383} (1996) 44,
hep-th/9512059;
N. Seiberg, Phys. Lett. B {\bf 390} (1996) 169, hep-th/9609161;
N. Seiberg and E. Witten, Nucl. Phys. B{\bf 471} (1996) 121,
 hep-th/9603003.}

\bigskip
\Title{\vbox{\baselineskip12pt
\hbox{hep-th/9710014}}}
{\vbox{\centerline{\bf{ SEMICLASSICAL DECAY OF  }}
\vskip2pt\centerline{\bf{NEAR-EXTREMAL FIVEBRANES }}}}

{
\baselineskip=11pt
\bigskip
\centerline{ 
Juan Maldacena and Andrew Strominger }
\bigskip
\centerline{\sl Department of Physics}
\centerline{\sl Harvard University }
\centerline{\sl Cambridge, MA 02138}

\bigskip
\centerline{\bf Abstract}
{We argue that a near-extremal charge-$k$ type II 
NS fivebrane can be reliably described in semiclassical string 
perturbation theory as long as both $k$ and $\mu \over k$
are large, where $\mu$ is the energy density in string units. For 
a small value of the asymptotic string coupling $g$, the dynamics in the throat surrounding the fivebrane
reduces to the CGHS model with massive fields. We find that the energy  
density leaks off the brane in the form of Hawking radiation 
at a rate of order $1 \over k^{7/2}$ in string units independently of $g$ 
to leading order.  In the $g\to 0$ limit the radiation persists but never 
reaches asymptotic infinity because the throat becomes infinitely long.}
\Date{}

One of the most surprising and intriguing results of recent years is the discovery 
of hitherto-unsuspected quantum field theories without gravity 
in six \sixr\ and five \fiver\ dimensions. These theories are
incompletely understood and have been analyzed by a variety of methods. 
In this note we shall adopt the method of \jmf\ and 
deduce some properties of the fivebrane 
theories \refs{\sixr,\nte} 
from the corresponding near-extremal black fivebrane solution.
In particular we shall focus, following \nte, on the theory of $k$ 
nearby fivebranes 
with finite energy densities in string units in the limit 
in which the asymptotic value of the string coupling $g \to 0$. 

The near-extremal string-frame charge-$k$ fivebrane solution is \ghas\
\eqn\dss{ds^2= -(1-{r_0^2 \over r^2})dt^2 +(1+{k\apm \over r^2})
({dr^2\over 1-{r_0^2\over r^2}}+r^2d\Omega_3^2)+dy_5^2,}
\eqn\hhj{e^{2\phi}=g^2(1+{k\apm\over r^2}).}
This configuration has string-frame energy per unit five-volume  
\eqn\mfv{{M \over V_5}={M_s^6 \over (2 \pi)^5 }
({k \over g^2} +\mu),}
where
\eqn\mq{\mu ={r_0^2 M_s^2\over g^2}.}
$\mu/(2\pi)^5 $ is the dimensionless energy density in string units 
and $M_s=1/\sqrt\apm$. 
As in \nte\ we wish to consider the case 
that $\mu$ is order one and so 
$r_0\sim g\sqrt{\apm}$. To analyze the limit $g\to 0$ it 
is useful to introduce coordinates
\eqn\rf{r=r_0cosh\sigma.}
In these coordinates the horizon is at $\sigma=0$, and as 
$g \to 0$ the asymptotically flat region moves off to infinity. For $g\to 0$
we are left 
with the two-dimensional black hole \ewm\
\eqn\cg{ds^2=- tanh^2 \s dt^2+k\apm {d\s^2 }+k\apm d\Omega_3^2
+dy_5^2.}
\cg\ is independent 
of $\mu$. The $\mu$-dependent dilaton is 
\eqn\dlt{e^{2\phi}={k \over \mu cosh ^2\s} .}
Notice that the value of $\phi$ at the horizon (located at $\sigma =0$)
 is independent of 
$g$ when expressed in terms of $k$ and $\mu$. 
The dynamics of this theory were studied in \cghs. In fact the CGHS model was 
originally derived by considering precisely this limit of near-extremal NS fivebranes. 

String loop perturbation theory is good if 
\eqn\km{{k \over \mu}\ll 1}
so that the dilaton at the horizon is small.
$\apm$ perturbation theory is good if the curvatures are small
\eqn\agd{k \gg 1.}
If both expansions are good then
\eqn\bg{\mu \gg k\gg 1.}

It is evidently possible to have both expansions good for $g\to 0$.
In this case one can compute the leakage of energy off the brane 
as semiclassical Hawking radiation \foot{One may also reliably 
compute the Bekenstein-Hawking entropy density as $S= \mu \sqrt{k}/(2\pi)^4$. 
It is of interest to note in the present context that this formula 
was microscopically reproduced in \jmf\ as the entropy of a gas of 
strings on the fivebrane with tensions $1/2\pi k \apm$ and central charge 
6.}. The Hawking temperature 
is 
\eqn\th{T_H={1 \over 2 \pi \sqrt{ k\apm }.}}
This is of order the mass gap for throat excitations. Hence 
there will be Hawking radiation of string states, even 
accounting for greybody factors. Angular momentum
produces a contribution to the  
effective mass proportional to ${\ell \over \sqrt {k \apm}}$ 
along the throat,
so higher angular modes can also be emitted at the temperature \th .
The rate of mass loss is 
\eqn\dmdtnc{{d\mu \over dt}\sim -{M_s \over k^{7/2}}.}
In the context of two dimensional black holes the rate of mass loss
was estimated in 
\gs\ as 
\eqn\dmdt{{d\mu \over dt}\sim -{M_s \over k}.}
The difference is due to the fact that this last calculation 
does not take into account the emission of gravitons with momenta
along the  directions parallel of the fivebrane\foot{
We thank S. Gubser for pointing this out to us. }.  
 So \dmdt\ is 
correct  when the fivebrane is compactified 
on a torus with sizes much smaller than $\sqrt{k \alpha'} $ 
(and bigger than $\sqrt{ \alpha'/k}$).  
There is also some mass loss due to fivebrane emission which 
is finite for $g\to 0$ but suppressed in the regime \bg.

This suggests that the theory of $k$ NS fivebranes does not 
decouple from the throat theory for $g\to 0$ for energies of order 
$M_s$, although it does decouple from 
the asymptotically flat region. It may nevertheless 
decouple in the $k\to \infty$ 
limit needed for matrix 
theory \sfs.  We also note that, for finite $k$ and $g \to 0$, 
there is no Hawking 
radiation for excitation energies far below $M_s$ because of the mass gap in 
the throat. Hence the excitations of the 5+1 conformal field theories 
at the low energy fixed point \sixr\ should decouple from the throat. 

In the matrix description of $k$ type II fivebranes 
\refs{ \db \verlinde \mm -\ewch}
 the throat 
and asymptotically flat regions are the Coulomb branch of a 
two-dimensional gauge theory, while the branes themselves are 
described by the Higgs branch. Our results suggest that 
excitations of the Higgs branch with energy of order $M_s$ 
can leak on to the Coulomb branch even for $g\to 0$.

 It would be of interest to reconcile our observations with 
those of  \refs{\nte, \mm,\ewch}.

\centerline{\bf Acknowledgements}
We are grateful to N. Seiberg and E. Witten for useful discussions.
This work was supported in part by DOE grant DE-FG02-96ER40559.
\listrefs
\bye